\pdfoutput=1
\documentclass[aps,prl,twocolumn,superscriptaddress]{revtex4-2}
\usepackage{graphicx}
\usepackage{siunitx}
\usepackage{amsmath}
\usepackage{amsfonts}
\usepackage{physics}
\usepackage{ulem}
\usepackage[colorlinks=true,
            linkcolor=blue,   
            citecolor=blue,   
            urlcolor=blue,    
            filecolor=blue]{hyperref}
\usepackage{xcolor}
\usepackage{upgreek} 
\usepackage{orcidlink}
\usepackage{braket}

\begin{document}

\title{Robust Two-Qubit Geometric Phase Gates Using Amplitude and Frequency Ramping}
\preprint{LLNL-JRNL-2013345}

\author{C. M. Bowers\,\orcidlink{0009-0005-3812-832X}}
\email{christina.bowers@colorado.edu}
\affiliation{National Institute of Standards and Technology, 325 Broadway, Boulder, Colorado 80305, USA}
\affiliation{University of Colorado, Boulder, Colorado 80309, USA}

\author{D. Palani\,\orcidlink{0000-0002-0938-9273}}
\affiliation{National Institute of Standards and Technology, 325 Broadway, Boulder, Colorado 80305, USA}
\affiliation{University of Colorado, Denver, Colorado 80204, USA}

\author{J. J. Barta\,\orcidlink{0009-0005-2142-2127}}
\affiliation{National Institute of Standards and Technology, 325 Broadway, Boulder, Colorado 80305, USA}
\affiliation{University of Colorado, Boulder, Colorado 80309, USA}

\author{T. H. Guglielmo\,\orcidlink{0000-0003-1024-9304}}
\affiliation{Lawrence Livermore National Laboratory, Livermore, California 94550, USA}

\author{S. B. Libby\,\orcidlink{0000-0001-5811-2273}}
\affiliation{Lawrence Livermore National Laboratory, Livermore, California 94550, USA}

\author{D. Leibfried\,\orcidlink{0000-0001-8442-625X}}
\affiliation{National Institute of Standards and Technology, 325 Broadway, Boulder, Colorado 80305, USA}

\author{D. H. Slichter\,\orcidlink{0000-0002-1228-0631}}
\email{daniel.slichter@nist.gov}
\affiliation{National Institute of Standards and Technology, 325 Broadway, Boulder, Colorado 80305, USA}

\begin{abstract}
We demonstrate a method for generating entanglement between trapped atomic ions based on adiabatically ramped state-dependent forces. By ramping both the amplitude of the state-dependent force and the motional mode frequencies, we realize an entangling operation that is robust to motional mode occupation and drifts in the mode frequencies. We measure Bell state fidelities above 0.99 across a broad range of ramp parameters and with motional occupations up to 10 phonons. This technique enables high-fidelity entangling operations without ground-state cooling, has a reduced calibration overhead, and is well suited for both quantum logic spectroscopy applications and scalable quantum computing architectures.
\end{abstract}

\maketitle

The controlled creation of high-fidelity entanglement is crucial for quantum applications across all physical platforms. It is valuable to make this process robust to experimental imperfections or miscalibrations, especially in large-scale quantum systems. Trapped atomic ions possess highly coherent and precisely controllable internal electronic and nuclear degrees of freedom~\cite{Marshall2025, Smith2025, Wang2021, Loschnauer2025}.
They are typically entangled by coupling these internal (``spin'') states to the ions' shared motional modes, which can be treated as quantum harmonic oscillators when appropriately cooled.  The earliest proposed trapped ion entangling gate required that the motional mode be prepared in its ground state, with any motional excitation directly manifesting as entangling gate infidelity~\cite{Cirac1995}. Alternatively, far-off-resonant, state-dependent excitation of the ion motion via static state-dependent forces or instantaneous ``kicks'' can provide the desired entanglement with reduced sensitivity to the state or coherence properties of the motional mode~\cite{Cirac2000, Mintert2001,Calarco2001,Sasura2003,Poulsen2010, Garcia-Ripoll2003, Mehdi2021}. 

Currently, geometric phase gates are the most widely used motional-state-insensitive entangling schemes for trapped ions~\cite{Milburn1999, Solano1999, MS1999, MS2000, Milburn2000}. Here, the shared ion motion is coherently displaced along a closed phase-space trajectory by a force that depends on the internal states, with each internal state acquiring a geometric phase according to the circumscribed area in phase space. Crucially, the motion must return to its initial state at the end of the entangling operation to eliminate any entanglement between the internal and motional degrees of freedom. While the fidelity of these gates is nominally insensitive to motional mode occupation, various experimental imperfections can prevent perfect closure of motional trajectories, causing gate infidelities that scale with the mode occupation. While this may be tolerable for certain experimental parameters~\cite{Kirchmair2009, Srinivas2020, Barthel2023}, in general the highest fidelity demonstrations of geometric phase gates have been done with the motional modes cooled to the ground state, and with careful attention paid to reduce motional decoherence rates and calibrate the motional frequency precisely, including accounting for motional frequency drifts~\cite{Ballance2016, Gaebler2016, Srinivas2021, Clark2021, Loschnauer2025}. Motional state sensitivity can also be mitigated by modulating the amplitude, phase, and/or frequency of the state-dependent force driving the gate~\cite{Hayes2012, Choi2014, Green2015, Zarantonello2019,Leung2018, Bentley2020, Milne2020, Duwe2022, Kang2023, Sutherland2024, Ruzic2024, Ellert-Beck2025}, by driving state-dependent forces at multiple frequencies~\cite{Haddadfarshi2016, Shapira2018, Webb2018, Sutherland2020, Blumel2021, Blumel2021a, Orozco-Ruiz2025}, or through higher-order sidebands~\cite{Sameti2021, Shapira2023}. More generally, sensitivities to system parameters can be reduced with the introduction of adiabatic methods, as exemplified by stimulated Raman adiabatic passage (STIRAP)~\cite{Bergmann1998}. In large-scale quantum computing architectures, overhead from ground state cooling, cold ion transport, and associated calibrations increases operational complexity and consumes the majority of the time budget in current state-of-the-art quantum algorithms~\cite{Pino2021,Moses2023, Ransford2025}. Reducing the motional-state sensitivity of gate operations may help address these challenges.

\begin{figure*}[t!]
\includegraphics[width=\textwidth]{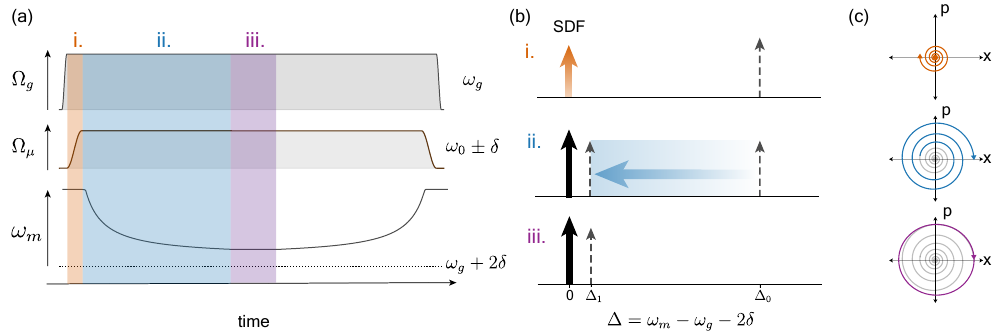}
\caption{
Gate pulse sequence and motional trajectories. 
Panel (a) shows the time-dependent amplitude envelopes $\Omega_g(t)$ of the magnetic gradient drive at $\omega_{g}$ (top) and $\Omega_\mu(t)$ of the bichromatic magnetic fields at $\omega_{0} \pm \delta$ (middle), as well as the frequency of the motional mode $\omega_m(t)$ (bottom). We plot the detuning $\Delta$ between the state-dependent force (thick arrow) and the motional mode (dotted arrow) in (b), and a cartoon of the motional phase space trajectory in (c), for the three highlighted regions in (a). We first ramp on $\Omega_g(t)$ in a time $\tau_g$. Next, in the orange region (i), $\Omega_\mu$ (and thus the SDF) is ramped on over $\tau_\mu$ at far-detuned $\Delta_0$. In the blue region (ii), the motional frequency is ramped from $\Delta_0$ to $\Delta_1$ over duration $\tau_m$. In the purple region (iii), the motional frequency is held constant at $\Delta_1$, followed by time-reversed versions of all ramps to complete the sequence.}
\label{fig:pulseseq}
\end{figure*}

In this work, we demonstrate an entangling gate for trapped ions whose fidelity is robust both to motional occupation and to drifts or offsets of the motional frequency. We accomplish this by adiabatically ramping both the amplitude of the state-dependent force (SDF) and the frequency of the motional mode, suppressing errors due to residual spin–motion entanglement. We demonstrate Bell-state fidelities exceeding 0.99 that are independent of motional occupation for up to 10 phonons, well above the Doppler-cooled mode occupation. We also show that the fidelity is reduced by only $\approx1\%$ when the motional mode frequency is miscalibrated by an amount equal to the inverse of the gate duration. The performance of the gate is insensitive to the exact form of the ramps, as long as they are sufficiently adiabatic. Unlike typical implementations of geometric phase gates, this method does not require joint calibration of the gate duration and detuning between the SDF and motional mode frequency. We perform these entangling gates on two $^{40}$Ca$^+$ ions, but we use a driving scheme that is well-suited for use as a mixed-species entangling gate~\cite{Tan2015, Ballance2015, Bruzewicz2019a, Hughes2020} in quantum information processing or quantum logic spectroscopy~\cite{Schmidt2005}. The drive consists of a strong MHz-frequency oscillating magnetic field gradient combined with weaker microwave-frequency magnetic fields~\cite{Sutherland2019, Srinivas2021}. Adiabatically ramped gates can be used with any gate drive method, but lend themselves naturally to laser-free gate implementations because the increased duration due to ramping does not incur additional spontaneous emission error~\cite{Ozeri2007, Moore2023, Boguslawski2023}.

We consider the gate Hamiltonian for two qubits of frequency $\omega_0$ described by collective spin operators $\hat{S}_i = \hat{\sigma}_{i,1} + \hat{\sigma}_{i,2}$, where the $\hat\sigma_{i,k}$ are the Pauli operators ($i\in\{x,y,z\}$) for ion $k$, and a motional mode with frequency $\omega_m$ and ladder operators $\hat{a}$ and $\hat{a}^\dagger$. We apply a magnetic field gradient oscillating at $\omega_g$ (with $\omega_g\sim\omega_m\ll\omega_0$) with strength $\Omega_g$ and two homogeneous magnetic fields oscillating at $\omega_0\pm\delta$ (with $\delta\ll\omega_0$) with strength $\Omega_\mu$~\cite{Sutherland2019, Srinivas2021}. Transforming into the interaction picture with respect to $\hbar\omega_0\hat{S}_{z} + \hbar\omega_m\hat{a}^\dagger\hat{a}$, which we call the ``ion frame'', and neglecting fast-rotating terms near $2\omega_0$, gives the gate Hamiltonian~\cite{Sutherland2019} 
\begin{equation}
\begin{aligned}
\label{ionframe}
    H_{\text{ion}} &= 2\hbar\Omega_\mu \hat{S}_x \cos(\delta t) \\
    &+ 2\hbar\Omega_g \cos(\omega_gt) \hat{S}_z \left[\hat{a}e^{-i \omega_m t} + \hat{a}^\dagger e^{i \omega_m t}\right]\, .
\end{aligned}
\end{equation}
The first (``bichromatic'') term drives Rabi oscillations of the qubits with time-dependent Rabi frequency $\Omega_{\mu}\cos(\delta t)$, while the second (``gradient'') term describes a state-dependent force oscillating at $\omega_g$. Because the two terms do not commute, however, the state-dependent force is modulated by the bichromatic drive. Transforming into the interaction picture with respect to the first term (the ``bichromatic'' interaction picture) modifies the state-dependent force, giving~\cite{Roos2008, Sutherland2019}
\begin{equation}
\label{eq:fullbichrom}
\begin{aligned}
\hat{H}_I(t) &= 2\hbar \Omega_g \cos(\omega_g t) 
    \left( \hat{a} e^{-i \omega_m t} + \hat{a}^\dagger e^{i \omega_m t} \right)
    \Big[
        \hat{S}_z
            J_0\!\left( \frac{4 \Omega_\mu}{\delta} \right) \\
            &+ 2 \hat{S}_z \sum_{n=1}^{\infty} 
                J_{2n}\!\left( \frac{4 \Omega_\mu}{\delta} \right)
                \cos(2 n \delta t) \\
        &+ 2 \hat{S}_y 
        \sum_{n=1}^{\infty}
            J_{2n-1}\!\left( \frac{4 \Omega_\mu}{\delta} \right)
            \sin\!\big([2n - 1] \delta t\big)\Big],
\end{aligned}
\end{equation}
where $J_k$ is the $k$th Bessel function of the first kind. The modulation of the state-dependent force at $\omega_g$ by the non-commuting bichromatic qubit drive spreads the total force over multiple frequencies $\omega_g\pm m\delta$ for integers $m$, producing either $\hat{S}_y$ (odd $m$) or $\hat{S}_z$ (even $m$) interactions. The frame transformation operator between this bichromatic interaction picture and the ion frame is the identity operator as long as the bichromatic term is ramped on and off slowly compared to $1/\delta$~\cite{Sutherland2019}. In this work, we consider the case where $\Omega_g$, $\Omega_\mu$, and $\omega_m$ are time-dependent and are ramped with durations $\tau_{g}$, $\tau_\mu$, and $\tau_m$, respectively. We choose $\delta$ such that the $J_2$ term is near resonant with $\omega_m$ and neglect the other terms, whose impact we consider later. The resulting gate Hamiltonian in the bichromatic interaction picture is then approximately~\cite{supplementalMaterial}

\begin{equation}
\label{eq:gateham}
\begin{aligned}
\hat{H}_I(t) &\approx 4\hbar \cos(\omega_g t) \cos(2\delta t)\Omega_g(t) J_{2}\!\left( \frac{4 \Omega_\mu(t)}{\delta} \right) \\
&\times \hat{S}_z\big( \hat{a} e^{-i \phi(t)} + \hat{a}^\dagger e^{i \phi(t)} \big),
\end{aligned}                
\end{equation}
where the phase $\phi(t) = \int\limits_0^t \omega_m(t')dt'$ replaces the linear phase accumulation $\omega_m t$ from Eq.~\eqref{eq:fullbichrom}. This Hamiltonian can be used to produce an effective $\hat{\sigma}_{z,1}\hat{\sigma}_{z,2}$ entangling interaction~\cite{Srinivas2021,supplementalMaterial}. The instantaneous detuning of this state-dependent force from the motional frequency is $\Delta(t)=\omega_m(t)-\omega_g-2\delta$. 

In Figure~\ref{fig:pulseseq}(a), we plot $\Omega_g(t)$, $\Omega_\mu(t)$, and $\omega_m(t)$, with the detunings $\Delta(t)$ and schematic motional phase space trajectories for one of the qubit states shown in Figs.~\ref{fig:pulseseq}(b) and (c), respectively. The gradient amplitude $\Omega_g(t)$ is ramped on first to its maximum value $\Omega_{g0}$, with $\tau_g$ chosen to minimize spectral content at $\omega_m$ to avoid off-resonant motional excitation. Next, the microwave magnetic field amplitude $\Omega_\mu(t)$ is ramped on to its maximum value $\Omega_{\mu0}$, with $\tau_\mu$ chosen to be adiabatic with respect to the detuning from the state-dependent force, $|\tau_\mu\Delta|\gg1$, which motivates ramping with the largest practical $|\Delta|$. Viewed in motional phase space in Fig.~\ref{fig:pulseseq}(c), adiabatically ramping $\Omega_{\mu}$ in this regime produces spiral trajectories out from the origin at the beginning of the gate pulse and into the origin at end of the gate pulse, which reduces errors due to residual spin-motion entanglement as described theoretically in Refs.~\cite{Leibfried2007, Tinkey2022, Sutherland2024, supplementalMaterial}. 
Finally, $\omega_m(t)$ is ramped such that $\Delta$ moves from a ``far-detuned" regime $|\Delta| = |\Delta_0| \gg \Omega_{g0} J_2(4\Omega_{\mu0}/\delta)$, where the microwave magnetic field is ramped on, to a ``near-detuned" one with $|\Delta|={|\Delta_1|\gtrsim\Omega_{g0}J_2(4\Omega_{\mu0}/\delta)}$. We require that $\Delta$ changes adiabatically, with $\left |\frac{1}{\Delta^2}\frac{d\Delta}{dt}\right |\ll 1$ and $\frac{d\Delta}{dt}=0$ at the start and end of the ramp. This provides the robustness of a gate operated in the far-detuned or ``weak-field'' regime~\cite{Cirac2000, Mintert2001, MS2000}, with the additional reduction in residual spin-motion entanglement afforded by amplitude ramping~\cite{Sutherland2024, supplementalMaterial}, but accumulates geometric phase much more rapidly. Any amplitude and frequency ramp shapes that meet these adiabaticity criteria and have appropriate $\Delta_0$ and $\Delta_1$ can be used (see End Matter), affording wide flexibility in implementation. Region (iii) is an optional ``flat-top'' regime ($\Omega_g$, $\Omega_\mu$, and $\omega_m$ are constant) whose duration can be adjusted to acquire the desired geometric phase. The ramps are time-reversed after the flat-top region. 

We trap two $^{40}\text{Ca}^+$ ions in a surface-electrode ion trap with current-carrying electrodes to produce the strong magnetic field gradient at $\omega_g$ and the magnetic fields at $\omega_0\pm\delta$, as described in~\cite{Srinivas2019, Srinivas2020, Srinivas2021, supplementalMaterial}. We use the ground state Zeeman sublevels $^2S_{1/2}\ket{m_J=-1/2}\equiv\ket{\downarrow}$ and $^2S_{1/2}\ket{m_J=1/2}\equiv\ket{\uparrow}$ as a qubit, with a splitting $\omega_0=2\pi\times596$ MHz set by the applied quantization magnetic field of $B_0\approx21.3$ mT. The gate is performed on a radial out-of-phase motional mode of the two ions with $\omega_{m}\approx2\pi\times6.8$ MHz, oriented at roughly 20 degrees from the electrode surface normal. Doppler cooling and fluorescence readout are performed using 397 nm laser light resonant with the ${^2S_{1/2}}\leftrightarrow{^2P_{1/2}}$ transition along with 866 nm laser light to repump the ${^2D_{3/2}}$ level. We use a narrow 729 nm laser to drive the ${^2S_{1/2}}\leftrightarrow{^2D_{5/2}}$ transition for resolved sideband cooling and population shelving prior to readout~\cite{Dehmelt1982, Diedrich1989, Roos1999}. We cool all four radial modes close to their ground states using electromagnetically induced transparency (EIT) cooling~\cite{Morigi2000, Roos2000a}, and both axial modes using resolved sidebands on the ${{^2S_{1/2}}\leftrightarrow{^2D_{5/2}}}$ transition. We set the fully-ramped-on value of $4\Omega_{\mu0}/\delta\approx2.405$, near the first zero crossing of $J_0$, to increase the qubit coherence through intrinsic dynamical decoupling (IDD), as this particular gate mechanism requires a magnetic-field-sensitive qubit~\cite{Sutherland2019, Srinivas2021, Srinivas2020}. However, the adiabatic ramping technique is general and can be used in gates on magnetic-field-insensitive qubits. The motional frequency is ramped by changing the amplitude of the trap rf drive and thus the strength of the pseudopotential confinement~\cite{Wineland1998}. The ramp function is chosen to maintain a constant adiabaticity parameter $\alpha\equiv\frac{1}{\Delta^2}\frac{d\Delta}{dt}$ when ramping from the far detuning $\Delta_0$ to the near detuning $\Delta_1$. Solving for $\Delta(t)$ gives the general form
\begin{equation}
\label{eq:adiabaticity}
    \Delta(t) = \frac{\Delta_0}{1 - \alpha \Delta_0 t}
\end{equation}
during the ramp, with the ramp duration fixed by the choice of $\alpha$ to be $\tau_{m}=\left|\frac{\Delta_0-\Delta_1}{\alpha\Delta_0\Delta_1}\right|$. We modify this ramp shape smoothly so the ramp approaches $\Delta_0$ and $\Delta_1$ with $\frac{d\Delta}{dt}=0$ (see End Matter). The entire gate operation consists of two identical pulse sequences of the type shown in Fig.~\ref{fig:pulseseq}(a) separated by a qubit $\pi_x$ pulse to implement a spin echo and Walsh modulation~\cite{Hayes2012}, preceded and followed by qubit $\pi_x/2$ pulses. This nominally takes the initial state $\ket{\uparrow\uparrow}$ to the Bell state $\frac{1}{\sqrt{2}}(\ket{\downarrow\downarrow} -i\ket{\uparrow\uparrow})$.

\begin{figure}
    \includegraphics[width=\columnwidth]{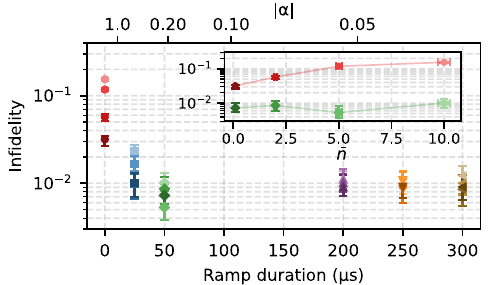}
    \caption{
    Bell state infidelity as a function of motional ramp duration and adiabaticity.
    Points for each ramp duration correspond to initial motional states with mean phonon occupations of $\bar{n} \approx 0, 2, 5,$ and~$10$, distinguished by progressively lighter shading to indicate increasing $\bar{n}$. The inset compares the unramped gate (red) with a $50~\mu\mathrm{s}$ ramp (green), illustrating the suppressed dependence of the ramped gate fidelity on the initial phonon occupation. Error bars correspond to 68\% confidence intervals, and may be smaller than plot markers.}
    \label{fig:rampdependence}
\end{figure}

To demonstrate the insensitivity of the ramped gate to motional mode occupation, we characterize Bell state infidelity as a function of $\tau_m$ for a range of mean phonon populations, as shown in Fig.~\ref{fig:rampdependence}. The magnitude of the adiabaticity parameter $|\alpha|$ is plotted on the top axis for each ramp duration. We prepare different motional populations using a resonant excitation of the mode whose phase is randomized from shot to shot relative to the SDF (see End Matter). This approach avoids heating the axial modes, which cause increased state readout errors when hot due to our laser beam geometry. The gate begins far detuned with $\Delta_0=2\pi\times153.5$ kHz, and the gradient and magnetic fields are ramped on sequentially with $\tau_g=10\, \mu$s and $\tau_\mu=30\,\mu$s with Blackman-Harris envelopes~\cite{Harris1978}. The motional frequency is then ramped with variable $\tau_m$ to $\Delta_1=2\pi\times15$ kHz and back to $\Delta_0$. For each $\tau_m$, we adjust the flat top duration to optimize Bell state fidelity, with zero ramp duration representing a gate with constant $\omega_m$ and $\Delta(t)=\Delta_1$. Bell state fidelity is calculated via population and parity analysis, with uncertainties estimated via nonparametric bootstrapping as in Ref.~\cite{Srinivas2021}. We do not correct for state preparation and measurement (SPAM) errors, which are negligible compared to the measured infidelities. The gate becomes progressively less sensitive to the initial average motional occupation $\bar{n}$ as the motional ramps become slower, reaching a roughly constant Bell state infidelity below $10^{-2}$ for ramp durations $\geq50~\mu\mathrm{s}$ ($|\alpha|\leq 0.2$) and $\bar{n}$ up to 10, with a best fidelity of $0.9947^{+0.0015}_{-0.0029}$ at a ramp duration of 50 $\mu$s and $\bar{n}$ of 5. Faster ramps with $|\alpha|>0.2$, as well as gates with no frequency ramping, show increasing infidelity with motional occupation. 

\begin{figure}[t]
    \includegraphics[width=\columnwidth]{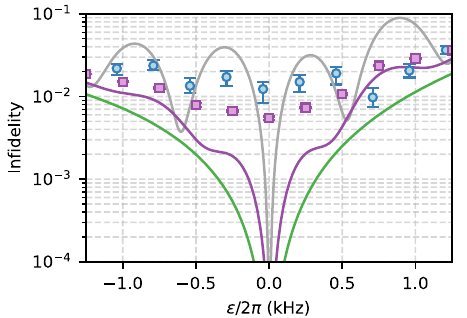}
    \caption{Bell state infidelity as a function of SDF detuning offset~$\varepsilon$ for ramped and non-ramped conditions. 
    Experimental data (blue circles) correspond to a ramped gate with parameters otherwise matching Fig.~\ref{fig:rampdependence} data with ramp duration $\tau_{m}=50~\mu\mathrm{s}$ on a ground-state-cooled ion. Numerical simulations of this gate that include qubit and motional dephasing are shown as purple squares. We also plot numerical simulations of an unramped gate (grey line) and gates with $\tau_{m}=50~\mu\mathrm{s}$ (purple line) and $\tau_{m}=300~\mu\mathrm{s}$ (green line), not including dephasing. All numerical simulations were performed in the ion frame and account for all four radial modes. Error bars correspond to 68\% confidence intervals, and may be smaller than plot markers.}
    
    \label{fig:detuning_offset}
\end{figure}

Single-loop non-ramped gates with fixed drive strength require a specific relationship between gate detuning and duration to avoid residual spin-motion entanglement, such that mode frequency fluctuations can result in gate errors~\cite{MS1999, MS2000}. Decoupling duration and detuning in a ramped gate gives robustness to miscalibrations or slow drifts of the mode frequency. It also simplifies calibration by enabling detuning and gate duration to be optimized independently rather than jointly. Figure~\ref{fig:detuning_offset} shows the measured Bell state infidelity for gates with a 50 $\mu$s motional ramp as a function of an offset $\varepsilon$ applied to the bichromatic detuning $\delta$, which mimics the effect of an offset or miscalibration of the motional frequency. The largest values of $\varepsilon$ shown are greater than the inverse of the total gate duration (see End Matter). Solid curves show numerical simulations (not accounting for qubit or motional dephasing), showing reduced sensitivity to such offsets for ramped gates compared to an unramped gate, with robustness improving for slower ramp rates.  
Numerically simulated data including independently calibrated motional and qubit dephasing rates quantitatively account for the gap between dephasing-free simulations and experimental data (see End Matter). If the other motional modes are not in their ground states during the gate, the cross-Kerr coupling between modes~\cite{Roos2008crosskerr, Nie2009} will introduce frequency shifts on the gate mode of a similar form to $\varepsilon$. The magnitude of the shift will depend on the specific confining potential~\cite{Ding2017}; we measure cross-Kerr shifts on the gate mode of $\lesssim30$ Hz per phonon in the spectator modes (see End Matter).

\begin{figure}[t]
    \includegraphics[width=\columnwidth]{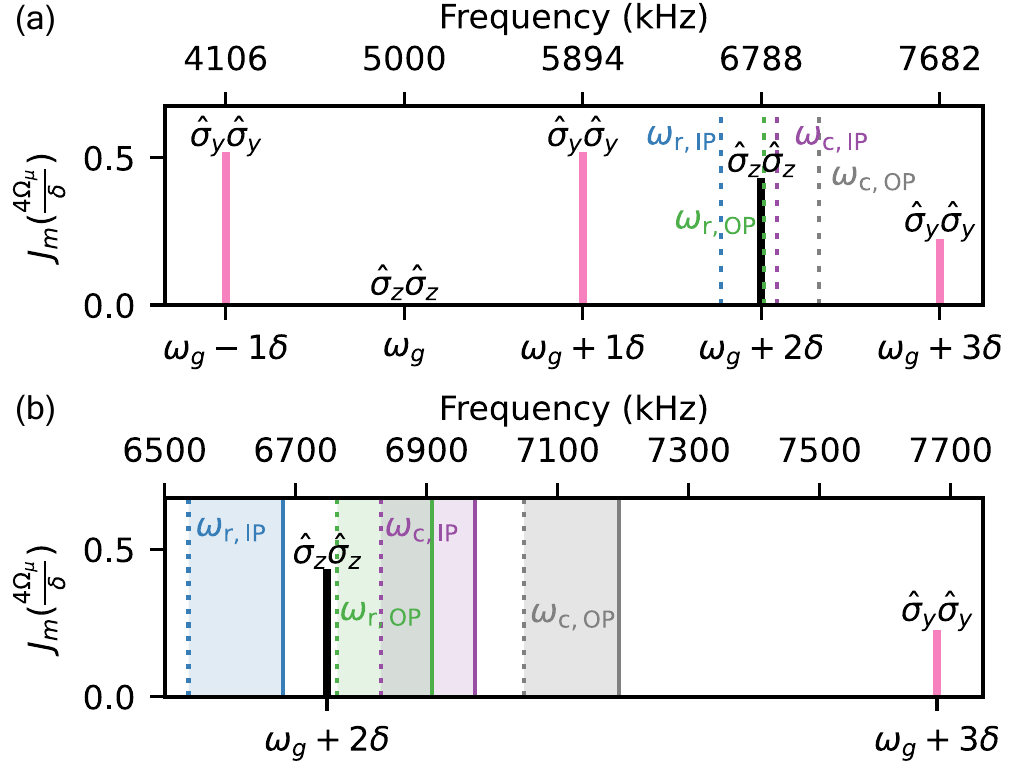}
    \caption{Spectrum of state-dependent forces in the bichromatic interaction picture. Panel (a) shows the frequencies and relative strengths of the different SDFs at the operating point (pink and black lines), along with the frequencies of the four radial modes at the near-detuned point (dotted lines). Panel (b) shows the range of frequencies spanned by each mode during the ramp, with the solid (dotted) lines indicating the far-detuned (near-detuned) limits of the ramp. }
    \label{fig:SDF}
\end{figure}

Because the two terms of the Hamiltonian in Eq.~\eqref{ionframe} do not commute, the effective state-dependent force in the bichromatic interaction picture is spread over multiple modulation sidebands at frequencies $\omega_g \pm m\delta$ for integer $m$, with a modulation index that depends on the strength and detuning of the bichromatic fields (see Eq.~\ref{eq:fullbichrom}). The odd and even sidebands produce effective $\hat{\sigma}_{y,1}\hat{\sigma}_{y,2}$ or $\hat{\sigma}_{z,1}\hat{\sigma}_{z,2}$ interactions, respectively. The SDF spectrum for our experimental parameters of $\omega_g=2\pi\times 5$ MHz, $\delta=2\pi\times 894$ kHz, and modulation index $4\Omega_{\mu0}/\delta\approx2.405$ is shown in Fig.~\ref{fig:SDF}(a), with the near-resonant SDF at $\omega_g+2\delta$ as in Eq.~(\ref{eq:gateham}). The frequencies $\omega_{c,\mathrm{IP}}$, $\omega_{r,\mathrm{IP}}$, $\omega_{c,\mathrm{OP}}$, and $\omega_{r,\mathrm{OP}}$ of the four motional modes that couple to the SDF\textemdash the in-phase ``center-of-mass'' and out-of-phase ``rocking'' modes for the in-plane (IP) and out-of-plane (OP) radial directions\textemdash are shown as dotted lines. All four modes acquire geometric phases, although the mode at $\omega_{r,\mathrm{OP}}$ is dominant; the ramping and detunings of all modes from the SDF help ensure closure of motional trajectories. However, the far-off-resonant $\hat{\sigma}_{y,1}\hat{\sigma}_{y,2}$ interactions do not commute with the $\hat{\sigma}_{z,1}\hat{\sigma}_{z,2}$ interaction, so their coupling to the four motional modes will limit the ultimate fidelity. These are not the limiting errors in our implementation, and it is possible to reduce the resulting infidelity below the $10^{-4}$ level with suitable $\delta$ and motional frequencies. Starting with larger $\Delta_0$ improves robustness and enables faster amplitude ramping, but the off-resonant $\hat{\sigma}_{y,1}\hat{\sigma}_{y,2}$ interactions and the requirement that no modes ramp through resonance with any SDF sidebands set practical limits on the maximum $\Delta_0$. Fig.~\ref{fig:SDF}(b) shows the frequency ranges over which the radial modes are ramped during the gate.

In summary, we demonstrate a high-fidelity, adiabatically ramped $\hat{\sigma}_z\hat{\sigma}_z$ entangling gate that does not require ground-state cooling and is robust to motional-frequency drifts and detuning errors. The ramped gate technique, which can be used with any type of geometric phase gate, relaxes technical requirements and could reduce time and calibration overhead for large-scale quantum systems based on ion shuttling, as well as for quantum logic spectroscopy. For multi-ion chains, the adiabatic elimination of spin–motion entanglement could relax the calibration requirements for simultaneous closure of multiple phase-space trajectories in geometric phase gates. The specific gate scheme we employ is experimentally advantageous for performing mixed-species entangling operations because the SDF can be produced from the same strong gradient plus only a weak bichromatic magnetic field pair for each species. Laser-free mixed-species quantum logic of this form could be used in an architecture where the ``data'' species does not require any laser light, and the ``helper'' species requires only low-power resonant lasers for Doppler cooling, state preparation, and readout. 

While performing experiments on amplitude-ramped gates, we became aware of related work on adiabatically ramped gates, subsequently reported in Ref.~\cite{Hughes2025}. Discussions with the authors of that work led us to add frequency ramping to our gate scheme.

\begin{acknowledgments}
\textit{Acknowledgments}---We acknowledge helpful discussions on amplitude and frequency ramped gates with R. T. Sutherland, who shared the key insight that adiabatically ramping a SDF to smaller detuning during the gate can maintain robustness while increasing speed. We acknowledge initial discussions on amplitude-ramped gates with R. T. Sutherland, D. Hayes, B. Bjork, S. Erickson, and P. Lee. We thank A. L. Carter and P. D. Kent for a careful reading of the manuscript. This work was supported in part by the NIST Quantum Information Program.  Part of this work was performed under the auspices of the U.S. Department of Energy by Lawrence Livermore National Laboratory under Contract DE-AC52-07NA27344. Support is also acknowledged from the U.S. Department of Energy, Office of Science, National Quantum Information Science Research Centers, Quantum Systems Accelerator. 

C.M.B., D.P., and D.H.S. conceptualized and planned the experiments. C.M.B. and D.P. carried out the experiments and analyzed the data, with assistance from J.J.B. and D.H.S.; C.M.B., T.H.G., and D.P. wrote the draft manuscript, and all authors participated in revisions. T.H.G. developed the analytical theory and carried out numerical simulations with assistance from C.M.B.; D.H.S., D.L., T.H.G., and S.B.L. secured funding for the work. D.H.S. supervised the work. 
\end{acknowledgments}

\textit{Data availability}---The data that support the findings of this article are openly available~\cite{DataAvailability}.

\normalem 
\bibliographystyle{apsrev4-2}
\providecommand{\noopsort}[1]{}\providecommand{\singleletter}[1]{#1}%

\newpage
\onecolumngrid 
\vspace{6pt}
\begin{center}
  \textbf{\large End Matter}
\end{center}
\vspace{6pt}
\twocolumngrid
\setcounter{equation}{0}
\renewcommand{\theequation}{A\arabic{equation}}

During the gate, the detuning $\Delta(t)=\omega_m(t)-\omega_g-2\delta$ is ramped such that its normalized rate of change remains constant and small:
\begin{equation}
\frac{1}{\Delta^2}\frac{d\Delta}{dt}=\alpha,\qquad|\alpha|\ll1\,.
\label{eq:adiabatic_condition}
\end{equation}
This defines $\alpha$, the adiabaticity parameter that describes the speed of the ramp.  Solving Eq.~(\ref{eq:adiabatic_condition}) for $\Delta(t)$ gives the functional form  
\begin{equation}
\frac{d\Delta}{dt}=\alpha\Delta^2
\quad\Rightarrow\quad
\Delta(t)=\frac{\Delta_0}{1-\alpha\Delta_0 t},
\label{eq:delta_solution}
\end{equation}
In this expression, $\Delta_0$ denotes the initial detuning. The ramp duration required to evolve from $\Delta_0$ to $\Delta_1$ follows from Eq.~(\ref{eq:delta_solution}) and is given by  
\begin{equation}
\tau_{m}=\frac{1}{\alpha}
\!\left(\!
\frac{1}{\Delta_0}-\frac{1}{\Delta_1}
\!\right)\!.
\label{eq:ramp_duration}
\end{equation}
Smaller absolute values of $\alpha$ produce slower ramps that better satisfy the adiabatic condition of Eq.~(\ref{eq:adiabatic_condition}).  In addition to this adiabatic condition, it is necessary to satisfy ${\tau_m\Delta_1^{3/2}\Delta_0^{-1/2}\gg1}$, as shown in the Supplementary Material~\cite{supplementalMaterial}, to minimize residual spin-motion entanglement at the end of the gate. We can rewrite this for our specific frequency ramp function using Eq.~(\ref{eq:ramp_duration}) as $\sqrt{\Delta_1/\Delta_0}\gg|\alpha|$. Practically speaking, this means that the near detuning $\Delta_1$ cannot be made too small while still achieving low residual spin-motion entanglement, and thus there is a limit on how much speedup can be achieved by ramping to smaller detuning during the gate. Once these criteria are met, the infidelity of the gate operation due to residual spin-motion entanglement is expected to scale with terms that go as the average motional occupation times $\mathcal{O}(\Omega_{g0}^2 \Delta_0^{-6}\tau_m^{-4})$, $\mathcal{O}(\Omega_{g0}^2 \Delta_0^{-6}\tau_m^{-2}\tau_\mu^{-2})$, and $\mathcal{O}(\Omega_{g0}^2 \Delta_0^{-6}\tau_\mu^{-4})$~\cite{supplementalMaterial}. These scalings can be made even more favorable, at the cost of additional ramp duration, by imposing further constraints that the higher time derivatives of the ramp functions are zero at their endpoints. The Supplementary Material provides an analytical derivation of these scalings and their underlying assumptions. 

Due to limitations in our microwave hardware, we choose to adjust our motional frequencies to ramp the detuning $\Delta$. One could alternatively ramp the SDF frequency (via the bichromatic detuning $\delta$), as in Ref.~\cite{Hughes2025}. To enable amplitude modulation of the trap rf drive to ramp the motional frequencies, the output of the signal generator producing the trap rf is split, with one arm going through an amplitude-stabilization circuit and the other going to a mixer. A dc-coupled arbitrary waveform generator (AWG) drives the IF port of the mixer to control the amplitude of the mixer output, which is then amplified and combined with the main amplitude-stabilized rf through the 20 dB forward port of a directional coupler. The resulting combined trap rf is amplified and sent to the trap rf resonator. 

The total waveform is shown in Fig.~\ref{fig:pulseseq}.  The experimental hardware is configured such that $\Delta(t)$ is proportional to the voltage output of an AWG, which parameterizes waveforms as a cubic spline~\cite{Bowler2013}. We adjust the spline knots near the start and end of the waveform, increasing the overall ramp duration slightly relative to Eq.~(\ref{eq:ramp_duration}), to enforce $\frac{d\Delta}{dt}=0$ at the ramp endpoints while maintaining $|\frac{1}{\Delta^2}\frac{d\Delta}{dt}|\leq|\alpha|$ at all times. The ramp durations given in the text include these start and end modifications. The ramp durations of $0$, $25$, $50$, $200$, $250$, and $300~\mu\mathrm{s}$ correspond to total durations of the pulse sequence shown in Fig.~\ref{fig:pulseseq}(a) of $468$, $505$, $527$, $698$, $718$, and $766~\upmu\mathrm{s}$, respectively. The corresponding total gate durations are then $937$, $1011$, $1055$, $1397$, $1437$, and $1533~\upmu\mathrm{s}$.

The dominant sources of infidelity in our demonstration are fast motional frequency dephasing, qubit dephasing due to magnetic field noise, and spurious qubit transitions due to resonant light leakage. These error mechanisms are not mitigated by the ramping techniques we employ and would limit the fidelity of any geometric phase gate\textemdash ramped or otherwise\textemdash at comparable gate durations.

We characterize motional dephasing by measuring the linewidth of the motional modes under weak resonant excitation by an electric field (“tickle”). After cooling near the ground state, we apply a weak, frequency-scanned tickle pulse and probe for the addition of a single motional quantum by attempting to drive a motion-subtracting sideband~\cite{McCormick2019}. The narrowest linewidth observed for tickle pulses longer than $2~\mathrm{ms}$ has a full width at half maximum of $\approx2\pi\times500~\mathrm{Hz}$, which we take as an upper bound on the motional dephasing rate. We also perform a ``motional Ramsey" experiment in the presence of the oscillating magnetic field gradient to determine a lower bound on the motional dephasing rate during the gate ~\cite{McCormick2019}. Dephasing is enhanced by the gradient drive, and we measure a time constant lower bound of $\approx1\,$ms for the exponential decay of motional coherence.

Qubit coherence is measured with Ramsey and spin-echo sequences, both with and without IDD~\cite{Sutherland2019,Srinivas2021}. We observe $1/e$ fringe contrast decay times of $T_2^\mathrm{R}\approx600\,\mu$s (Ramsey without IDD), $T_2^\mathrm{E}\approx1\,$ms (spin echo without IDD), and $T_2^\mathrm{E, IDD}\approx40\,$ms (spin echo with IDD). The envelopes of the fringe decay are approximately Gaussian in time with some non-monotonic structure, indicating the presence of slow magnetic field drifts as well as line noise and harmonics.

Leakage of resonant 397 nm light due to finite extinction of the acousto-optic modulators causes spurious transitions between the qubit states during the gate. We measured the probability of such transitions to be $\approx1.3\times10^{-3}$ at our maximum gate duration of 1.6 ms. This error channel was not included in the numerical simulations in Fig.~\ref{fig:detuning_offset}.

Gate errors due to non-adiabaticity of the motional ramp, motional mode drifts over the duration of the gate, off-resonant spectator mode interactions, and motional heating are all affected by the ramping technique. In general, the ramping technique reduces infidelity due to motional occupation (Fig. ~\ref{fig:rampdependence}), motional mode drifts (Fig. ~\ref{fig:detuning_offset}), and motional heating when compared to an unramped gate. In 
the absence of these error channels, we expect fidelity to be limited by the adiabaticity of the ramp and off-resonant spectator mode interactions.
We characterize approximate infidelity contributions from qubit dephasing, motional dephasing, motional heating, and off-resonant $\hat{\sigma}_y\hat{\sigma}_y$ interactions for the ground-state cooled $\tau_{m}=50~\mu\mathrm{s}$ gate using numerical simulations. We find that the estimated infidelities due to qubit dephasing and off-resonant interactions are $<2\times10^{-4}$, the infidelity from our measured heating rate of 1.8 quanta/second is $\approx1.5\times10^{-5}$, and the infidelity due to motional dephasing is $\approx5.4\times10^{-3}$.

For each Bell state fidelity shown, we perform 2,200 individual experimental trials of the entangling operation to analyze the population in the $\ket{\downarrow\downarrow}$, $\{\ket{\uparrow\downarrow},\ket{\downarrow\uparrow}\}$, and $\ket{\uparrow\uparrow}$ subspaces. Additionally, we perform 5,600 individual experimental trials of the same experiment followed by a qubit $\pi/2$ pulse with 28 different phase values from which we determine the average parity of the resulting state versus phase~\cite{Sackett2000}. The phases are distributed non-uniformly, bunched around values where the parity is near $\pm1$.

We prepare different motional occupations for the data in Fig.~\ref{fig:rampdependence} as follows. All modes of the ion crystal are cooled near their ground states using EIT cooling (radial modes) and resolved sideband cooling (axial modes). Controlled coherent excitation of the gate mode is then driven by applying a resonant potential to a trap electrode. Because this coherent drive is not phase-synchronized to the state-dependent force, each experimental shot realizes a coherent state of motion with complex amplitude $\xi = \abs{\xi}e^{i\theta}$ and randomized phase $\theta$, where we calibrate $|\xi|$ independently. We can then treat the ensemble motional state as a mixed state with density matrix 
\begin{align}
\rho(\xi) &= e^{-\abs{\xi}^2}\sum_{n=0}^\infty \frac{\abs{\xi}^{2n}}{n!}\ket{n}\bra{n} \\
&= e^{-\bar{n}} \sum_{n=0}^\infty \frac{\bar{n}^n}{n!}\ket{n}\bra{n}
\end{align}
By varying the drive duration, coherent states with different mean phonon occupations $\bar{n}=\mathrm{tr}(\rho \hat{a}^\dagger\hat{a})$ are prepared, corresponding to $\bar{n}=2.01(6)$, $5.0(1)$, and $10.0(3)$ for the data presented.
Rather than the phase-randomized coherent states used in this work, higher-$\bar{n}$ thermal states could be generated by controlled heating using Doppler beams or calibrated voltage noise applied to a trap electrode. In our apparatus, the elevated motional occupation of the axial modes after Doppler heating degraded the fidelity of our shelving pulses to the $D_{5/2}$ state such that state readout errors were much larger than the gate infidelity. The randomized coherent displacement approach avoids this issue by exciting only the gate mode. This is a suitable approach because infidelity due to residual spin-motion entanglement is linear in $n$~\cite{Sutherland2022, supplementalMaterial}, and therefore the total infidelity averaged over Fock states will only depend on the mean occupation $\bar{n}$ for both thermal states and phase-randomized coherent states.

Cross-Kerr coupling arises from anharmonic terms in the trapping potential and introduces mode–mode frequency shifts that depend on mode occupations. These couplings are described by the Hamiltonian $\hat{H}_{\mathrm{Kerr}}=\chi_{a,b}\hat{n}_a\hat{n}_b$, where $\chi_{a,b}$ denotes the coupling strength between modes $a$ and $b$ and the $\hat{n}$ are number operators for the modes~\cite{Nie2009,Roos2008crosskerr,Ding2017}. We measured these couplings by performing mode frequency spectroscopy with varying coherent excitation of coupled motional modes. For the present trap configuration, the measured couplings are $\chi_{xs,zr}=-28(2)~\mathrm{Hz}/\mathrm{phonon}$, $\chi_{xs,yr}=-41(3)~\mathrm{Hz}/\mathrm{phonon}$, and $\chi_{yr,zr}=9(1)~\mathrm{Hz}/\mathrm{phonon}$, where $xs$ represents the axial stretch mode and $yr$ and $zr$ correspond to the radial rocking modes aligned predominantly along the $y$ (in-plane) and $z$ (out-of-plane) radial axes. The $zr$ mode is the gate mode (at frequency $\omega_{r,\mathrm{OP}}$).

\end{document}


\title{Supplementary Material for ``Robust Two-Qubit Geometric Phase Gates using Amplitude and Frequency Ramping"}
\preprint{LLNL-JRNL-2013345}
\author{C. M. Bowers}
\email{christina.bowers@colorado.edu}
\affiliation{National Institute of Standards and Technology, 325 Broadway, Boulder, Colorado 80305, USA}
\affiliation{University of Colorado, Boulder, Colorado 80309, USA}

\author{D. Palani}
\affiliation{National Institute of Standards and Technology, 325 Broadway, Boulder, Colorado 80305, USA}
\affiliation{University of Colorado, Denver, Colorado 80204, USA}

\author{J. J. Barta}
\affiliation{National Institute of Standards and Technology, 325 Broadway, Boulder, Colorado 80305, USA}
\affiliation{University of Colorado, Boulder, Colorado 80309, USA}

\author{T. H. Guglielmo}
\affiliation{Lawrence Livermore National Laboratory, Livermore, California 94550, USA}

\author{S. B. Libby}
\affiliation{Lawrence Livermore National Laboratory, Livermore, California 94550, USA}

\author{D. Leibfried}
\affiliation{National Institute of Standards and Technology, 325 Broadway, Boulder, Colorado 80305, USA}

\author{D. H. Slichter}
\email{daniel.slichter@nist.gov}
\affiliation{National Institute of Standards and Technology, 325 Broadway, Boulder, Colorado 80305, USA}

\maketitle

\section{Experimental details}

Experiments are performed in a cryogenic surface-electrode trap operated at $10~\mathrm{K}$, where an rf pseudopotential driven at $\Omega_{\mathrm{RF}}/2\pi\approx69~\mathrm{MHz}$ creates the radial confinement. Ions are loaded by photoionization of neutral calcium from a thermal oven, using $423$ and $375$ nm laser beams. The ions are held 30~$\mu$m above the surface of the trap. Doppler cooling and fluorescence readout are performed on the $^2S_{1/2}\!\leftrightarrow\!^2P_{1/2}$ transition using co-propagating $397~\mathrm{nm}$ laser beams with $\sigma^{+}$ and $\sigma^{-}$ polarization, along with $866~\mathrm{nm}$ light to repump from the $^2D_{3/2}$ level. Qubit state initialization uses optical pumping into $\ket{\uparrow}$ using the 397 nm $\sigma^{+}$ laser beam. Shelving to the $^2D_{5/2}$ level for state readout is done with a narrow 729 nm laser, which is then depopulated after readout using 854 nm laser light. Readout errors are measured to be at the $10^{-4}$ level, primarily limited by occasional deshelving from $^2D_{5/2}$ and incomplete repumping from $^2D_{3/2}$. Near-ground-state cooling is achieved using two laser-based methods. The axial center-of-mass and stretch modes at $\omega_{\mathrm{COM}}/2\pi=2.0~\mathrm{MHz}$ and $\omega_{\mathrm{STR}}/2\pi=3.3~\mathrm{MHz}$ are sideband cooled~\cite{Wineland1998} on the $729~\mathrm{nm}$ transition, while the radial modes are cooled by electromagnetically induced transparency (EIT)~\cite{Morigi2000, Roos2000a} using a tripod configuration involving the two Zeeman sublevels of $S_{1/2}$ coupled to the $^2P_{1/2}\ket{m_J=1/2}$ state. The Zeeman splitting of the $^2P_{1/2}$ state is 197 MHz, so the $^2P_{1/2}\ket{m_J=-1/2}$ state is far detuned and can be neglected. The EIT scheme employs a strong $\sigma^{+}$-polarized beam blue-detuned by 70 MHz from ${^2S_{1/2}}\ket{m_J=-1/2}\leftrightarrow{^2P_{1/2}}\ket{m_J=1/2}$ and a weak $\pi$-polarized probe beam, with the 866 nm repump beam frequency tuned to optimize cooling. EIT cooling exhibits a $\approx1$ MHz cooling bandwidth across the $6.5$–$7.1~\mathrm{MHz}$ range, enabling simultaneous near-ground-state cooling of all four radial modes to $\bar{n}\approx0.1$.
The magnetic field gradient oscillating at $\omega_g/2\pi=5~\mathrm{MHz}$ is generated by three independent phase-synchronized channels of an arbitrary waveform generator (AWG)~\cite{Bowler2013,Srinivas2019,Srinivas2020}, each driving a current-carrying electrode parallel to the rf electrodes. The relative amplitudes and phases of the currents are tuned to cancel the 5 MHz magnetic field at the ion positions while maximizing the gradient strength~\cite{Srinivas2020}. The bichromatic magnetic field drive is generated by amplitude-modulated direct digital synthesizer sources that are amplified and combined on a single current-carrying electrode. Details of the drive electronics are given in Refs.~\cite{Srinivas2020, Knaack2024}. 

The parameters $\Omega_g(t)$ and $\Omega_\mu(t)$ used to characterize the strength of the state-dependent force and the bichromatic magnetic field can be expressed as~\cite{Srinivas2019, Srinivas2021}:
\begin{equation}
\begin{aligned}
\Omega_{g} &\equiv 
\frac{r_{0}\,(\hat{r}\!\cdot\!\nabla B_{g}(t))}{4}
\left.
\frac{d\omega_{0}}{dB_{z}}
\right|_{B_{z}={B}_{0}}
\\
\Omega_{\mu}&\equiv -\frac{B_x(t)}{2\hbar}\bra{\downarrow}\mu_x\ket{\uparrow}\,\,.
\end{aligned}
\end{equation}
Here $\nabla B_g(t)$ is the time-dependent amplitude of the gradient of the component of the magnetic field oscillating at $\omega_g$ along the quantization axis, $B_z$ is the component of the total magnetic field along the quantization axis, $B_0$ is the applied quantization field, $r_0=\sqrt{\hbar/(2M\omega_m)}$ is the ground-state extent of the wavefunction for the motional mode along the $\hat{r}$ direction with frequency $\omega_m$ for an ion of mass $M$, $B_x(t)$ is the time-dependent amplitude of the magnetic field perpendicular to the quantization axis oscillating near $\omega_0$ and $\mu_x$ is the component of the ion's magnetic moment pointed in the same direction.

\section{Derivation of the Bichromatic Hamiltonian}
Here we derive the full Hamiltonian of the experimental system described in the main body of the paper, with two ions and four radial modes.  We further show how the Hamiltonian can be transformed into the bichromatic interaction picture with time-dependent $\Omega_\mu(t)$ (microwave strength) and $\omega_m(t)$ (motional frequency of the gate mode).  Our starting point is the lab frame Hamiltonian for two ions interacting with a bichromatic microwave magnetic field, driven at two frequencies detuned by $\pm\delta$ (taking $\delta$ to be positive) from the qubit frequency $\omega_0$, and a MHz-frequency magnetic gradient oscillating at $\omega_g$.  We include a sum over all four radial motional modes, one out-of-phase or STR (s) and one in-phase or COM (c) mode for each radial direction, with the radial directions indexed by $i$. For this theory supplement we will work with $\hbar=1$ throughout, unless explicitly stated otherwise. We ignore the axial modes because there is no magnetic gradient along the axial modes due to the geometry of our current-carrying wires. The lab frame Hamiltonian is
\begin{align}
    H_{\text{lab}}(t) &= \frac12 \omega_0 \hat{S}_{z,c} + \sum_{i = 1}^2 \left[\omega_{i,c}(t) \hat{a}_{i,c}\dagg \hat{a}_{i,c} + \omega_{i,s}(t) \hat{a}_{i,s}\dagg \hat{a}_{i,s} \right]\nonumber\\
    &+ 2 \Omega_\mu(t) \hat{S}_{x,c} \left[\cos((\omega_0 + \delta) t) + \cos((\omega_0 - \delta)t) \right]\nonumber\\
    &+ 2\cos(\omega_g t) \sum_{i=1}^2 \Omega_{g,i}^{(c)}(t) \hat{S}_{z,c} \left(\hat{a}_{i,c} + \hat{a}_{i,c}\dagg\right) \nonumber\\
    &+ 2\cos(\omega_g t) \sum_{i=1}^2 \Omega_{g,i}^{(s)}(t) \hat{S}_{z,s} \left(\hat{a}_{i,s} + \hat{a}_{i,s}\dagg\right),
\end{align}
where $\hat{a}_{i,c}$ and $\hat{a}_{i,s}$ are the annihilation operators for the two COM modes and the two STR modes, respectively.  Each mode has an associated time-dependent motional frequency $\omega_{i,j}(t)$ and magnetic gradient Rabi rate $\Omega^{(j)}_{g,i}(t)$, where $i$ indexes the mode directions and $j$ indexes the modes along a given direction, defined as~\cite{Srinivas2019, Srinivas2021}
\begin{equation}
\Omega^{(j)}_{g,i}(t) \equiv 
\frac{r^{(j)}_{i0}\,(\hat{r}_i\!\cdot\!\nabla B_{g}(t))}{4}
\left.
\frac{d\omega_{0}}{dB_{z}}
\right|_{B_{z}={B}_{0}}.
\end{equation}
Here $\nabla B_g(t)$ is the time-dependent amplitude of the gradient of the component of the magnetic field oscillating at $\omega_g$ along the quantization axis, $B_z$ is the component of the total magnetic field along the quantization axis, $B_0$ is the applied quantization field, and $r^{(j)}_{i0}=\sqrt{\hbar/(2M\omega_{i,j})}$ is the ground-state extent of the wavefunction for the motional mode along the $\hat{r}_i$ direction with frequency $\omega_{i,j}$ for an ion of mass $M$ (we include $\hbar$ in this specific expression for clarity). The microwave field's Rabi rate $\Omega_\mu(t)$ is time dependent, taken to be the same for both ions, and defined as  
\begin{equation}
\Omega_{\mu}\equiv -\frac{B_x}{2\hbar}\bra{\downarrow}\mu_x\ket{\uparrow}\,\,.
\end{equation}
where $B_x$ is the magnetic field perpendicular to the quantization axis oscillating near $\omega_0$ and $\mu_x$ is the component of the ion's magnetic moment pointed in the same direction; again we have explicitly included $\hbar$ for clarity.

The spin operators are given by $\hat{S}_{i,c} = \hat{\sigma}_i \otimes \hat{\mathbb{I}} + \hat{\mathbb{I}}\otimes\hat{\sigma}_i$ and $\hat{S}_{i,s} = \hat{\sigma}_i \otimes \hat{\mathbb{I}} - \hat{\mathbb{I}}\otimes\hat{\sigma}_i$ where the $\{\hat{\sigma}_i\}$ are the Pauli matrices. The minus sign for $\hat{S}_{i,s}$ reflects the fact that the ions move out of phase with each other, but with equal amplitudes, in these modes. The time dependent Rabi rates and motional frequencies are defined as (for a single arm of the Walsh-1 sequence of length $t_\mathrm{arm}$, recalling that the total gate consists of two such arms with an intervening qubit $\pi$ pulse along with initial and final qubit $\pi/2$ pulses),
\begin{align}
    \Omega^{(c,s)}_{g,i}(t) = \Omega^{(c,s)}_{g0,i}
    &\begin{cases}
        e_{<g}(t) & 0 \le t \le \tau_g \\
        1 & \tau_g < t \le t_\mathrm{arm} - \tau_g\\
        e_{>g}(t) & t_\mathrm{arm} - \tau_g < t \le t_\mathrm{arm}
    \end{cases} \nonumber\\
    \Omega_\mu(t) = \Omega_{\mu0}
    &\begin{cases}
        0 & 0 \le t \le \tau_g \\
        e_{<\mu}(t) & \tau_g < t \le \tau_g + \tau_\mu \\
        1 & \tau_g + \tau_\mu < t \le t_\mathrm{arm} - \tau_g - \tau_\mu\\
        e_{>\mu}(t) & t_\mathrm{arm} - \tau_g - \tau_\mu < t \le t_\mathrm{arm} - \tau_g \\
        0 & t_\mathrm{arm} - \tau_g < t \le t_\mathrm{arm}
    \end{cases} \label{eq:envelopeFunctions}\\
    \omega_{i,j}(t) = \omega^{(i,j)}_0 + \omega^{(i,j)}_1
    &\begin{cases}
        1 & 0 \le t \le \tau_g + \tau_\mu \\
        e_{<m}(t) & \tau_g + \tau_\mu < t \le \tau_g + \tau_\mu + \tau_m \\
        0 & \tau_g + \tau_\mu + \tau_m < t \le t_\mathrm{arm} - \tau_g - \tau_\mu - \tau_m\\
        e_{>m}(t) & t_\mathrm{arm} - \tau_g - \tau_\mu - \tau_m < t \le t_\mathrm{arm} - \tau_g - \tau_\mu \\
        1 & t_\mathrm{arm} - \tau_g - \tau_\mu < t \le t_\mathrm{arm}
    \end{cases} \nonumber
\end{align}
where the $e_{<,>}$ are the specific envelopes for each ramp, and $\Omega^{(c,s)}_{g0,i}$, $\Omega_{\mu0}$, $\omega^{(i,j)}_0$, and $\omega^{(i,j)}_1$ are constants.  The gradient and microwave fields are ramped according to Blackman-Harris envelopes ($e_{<,>g,\mu}$)~\cite{Harris1978}, while the motional frequency ramp ($e_{<,>m}$) is done in such a way that the ramp speed remains adiabatic at all points with respect to the detuning (details of this ramp are provided in the End Matter of the main text).  

We now move into the ``ion frame'' with respect to the internal qubit and motional states.  This transformation operator is given by
\begin{widetext}
\begin{align}
    U_0(t) &= \exp\left(-\frac{i}{2} \omega_0t~ \hat{S}_{z,c}\right) \prod_{i=1}^2\exp{-i \hat{a}_{i,c}\dagg \hat{a}_{i,c} \intl_0^t \omega_{i,c}(t') dt'} \exp{-i \hat{a}_{i,s}\dagg \hat{a}_{i,s} \intl_0^t \omega_{i,s}(t') dt'} \\
    &\equiv \exp\left(-\frac{i}{2} \omega_0t~ \hat{S}_{z,c}\right) \prod_{i=1}^2\exp(-i \hat{a}_{i,c}\dagg \hat{a}_{i,c} \phi_{i,c}(t))\exp(-i \hat{a}_{i,s}\dagg \hat{a}_{i,s} \phi_{i,s}(t))\, ,
\end{align}
\end{widetext}
where we have defined the accumulated phase associated with each motional mode ${\phi_{i,j}(t)=\intl_0^t \omega_{i,j}(t') dt'}$. The lab frame Hamiltonian is then transformed into the ion frame Hamiltonian
\begin{align} 
    H_{\text{ion}}(t) &= U_0\dagg(t) H_{\text{lab}} U_0(t) + i \dot{U}_0 \dagg(t) U_0(t) \\
    &\approx 2 \Omega_\mu(t) \hat{S}_{x,c} \cos(\delta t) \nonumber\\
    &+ 2\cos(\omega_g t) \sum_{i=1}^2 \Omega_{g,i}^{(c)}(t) \hat{S}_{z,c} \left(\hat{a}_{i,c} e^{-i\phi_{i,c}(t)} + \hat{a}_{i,c}\dagg e^{i\phi_{i,c}(t)}\right) \nonumber\\
    &+ 2\cos(\omega_g t) \sum_{i=1}^2 \Omega_{g,i}^{(s)}(t) \hat{S}_{z,s} \left(\hat{a}_{i,s} e^{-i\phi_{i,s}(t)} + \hat{a}_{i,s}\dagg e^{i\phi_{i,s}(t)}\right)\label{eq:HamIonFrame}
\end{align}
where fast-oscillating terms near $2\omega_0$ have been dropped.  Numerical simulations reported in the main text are performed using the ion frame Hamiltonian Eq.~(\ref{eq:HamIonFrame}). We now go into the ``bichromatic frame'' with respect to the microwave Hamiltonian $H_\mu \equiv 2\Omega_\mu(t) \hat{S}_{x,c} \cos\delta t$.  This transformation operator is
\begin{align}
    U_\mu(t) &= \exp[-2i \hat{S}_{x,c} \intl_0^t \Omega_\mu(t') \cos(\delta t') dt'] \equiv \exp[-i \hat{S}_{x,c} F(t)].
\end{align}
Following the analysis in Ref.~\cite{Sutherland2019}, it can be shown that
\begin{widetext}
\begin{align}
    H_{\text{bi}}(t) &= U_\mu\dagg(t) H_{\text{ion}} U_\mu(t) + i \dot{U}_\mu \dagg(t) U_\mu(t) \\
    &= 2\cos(\omega_g t) \sum_{i=1}^2 \Omega_{g,i}^{(c)}(t) \left(\hat{a}_{i,c} e^{-i\phi_{i,c}(t)} + \hat{a}_{i,c}\dagg e^{i\phi_{i,c}(t)}\right) \left[\hat{S}_{z,c}\cos[2F(t)] + \hat{S}_{y,c}\sin[2F(t)]\right] \nonumber\\
    &+ 2\cos(\omega_g t) \sum_{i=1}^2 \Omega_{g,i}^{(s)}(t) \left(\hat{a}_{i,s} e^{-i\phi_{i,s}(t)} + \hat{a}_{i,s}\dagg e^{i\phi_{i,s}(t)}\right) \left[\hat{S}_{z,s}\cos[2F(t)] + \hat{S}_{y,s}\sin[2F(t)]\right] \label{eq:FullBichromaticHam}
\end{align}
\end{widetext}
The general expression for $F(t)$ is fairly complicated. However, the result can be simplified greatly because the bichromatic detuning $\delta$ is large compared to the inverse microwave ramp time, $\delta\tau_\mu \gg 1$. We can write $\Omega_\mu(t)=\Omega_{\mu0}h(t/\tau_\mu)$, where $h(t/\tau_\mu)$ is the piecewise-defined function in Eq.~(\ref{eq:envelopeFunctions}) with a scaled argument $\zeta\equiv t/\tau_\mu$. The function $h(\zeta)$ and all its derivatives with respect to $\zeta$ ($h^\prime(\zeta)$, $h^{\prime\prime}(\zeta)$, etc., where the prime notation means differentiation with respect to $\zeta$) will be dimensionless functions with values of order unity (or smaller). Differentiation of $h(\zeta)$ with respect to $t$ brings an additional factor of $\frac{d\zeta}{dt}=\tau_\mu^{-1}$ for each derivative, which we write separately from the ${h}^\prime$, ${h}^{\prime\prime}$, etc. We can then evaluate $F(t)$ by repeated rounds of integration by parts, leading to an infinite sequence of terms with increasing orders of $(\delta\tau_\mu)^{-1}$, the first few of which are shown in Eq.~(\ref{eq:FIntegralh2}) below:

\begin{widetext}
\begin{align}
    F(t) &= 2\Omega_{\mu0} \intl_0^{t} h(t'/\tau_\mu) \cos(\delta t') ~dt' \label{eq:FIntegralh}\\
    &= 2\Omega_{\mu0} \left [ \frac{h(t/\tau_\mu) \sin(\delta t)}{\delta} +\frac{1}{\delta\tau_\mu}\frac{ h^{\prime}(t/\tau_\mu) \cos(\delta t)}{\delta} - \frac{1}{(\delta\tau_\mu)^2}\frac{h^{\prime\prime}(t/\tau_\mu) \sin(\delta t)}{\delta} + \cdots \right ]  \label{eq:FIntegralh2} 
\end{align}
\end{widetext}
Because the derivatives $h^{\prime}(t/\tau_\mu)$, $h^{\prime\prime}(t/\tau_\mu)$, and so forth are all of order unity, and $(\delta\tau_\mu)^{-1}\ll 1$, we can truncate this series to lowest order to find
\begin{align}
    F(t) &\approx \frac{2\Omega_\mu(t) \sin(\delta t)}{\delta}.
\end{align}
With this simplified form, we can now perform the Jacobi-Anger expansion on Eq.~(\ref{eq:FullBichromaticHam}) to yield
\begin{widetext}
\begin{align}
    H_{\text{bi}}(t) &\approx 2\cos(\omega_g t) \sum_{i=1}^2 \Omega_{g,i}^{(c)}(t) \left(\hat{a}_{i,c} e^{-i\phi_{i,c}(t)} + \hat{a}_{i,c}\dagg e^{i\phi_{i,c}(t)}\right) \Bigg\{\hat{S}_{z,c}\left[J_0\left(\frac{4\Omega_\mu(t)}{\delta}\right) + 2\sum_{n=1}^\infty J_{2n}\left(\frac{4\Omega_\mu(t)}{\delta}\right)\cos (2n\delta t) \right] \nonumber \\
    &+ 2\hat{S}_{y,c}\sum_{n=1}^\infty J_{2n-1}\left(\frac{4\Omega_\mu(t)}{\delta}\right)\sin[(2n-1)\delta t] \Bigg\} \nonumber\\
    &+ 2\cos(\omega_g t) \sum_{i=1}^2 \Omega_{g,i}^{(s)}(t) \left(\hat{a}_{i,s} e^{-i\phi_{i,s}(t)} + \hat{a}_{i,s}\dagg e^{i\phi_{i,s}(t)}\right) \Bigg\{\hat{S}_{z,s}\left[J_0\left(\frac{4\Omega_\mu(t)}{\delta}\right) + 2\sum_{n=1}^\infty J_{2n}\left(\frac{4\Omega_\mu(t)}{\delta}\right)\cos (2n\delta t) \right] \nonumber \\
    &+ 2\hat{S}_{y,s}\sum_{n=1}^\infty J_{2n-1}\left(\frac{4\Omega_\mu(t)}{\delta}\right)\sin[(2n-1)\delta t] \Bigg\}\, , \label{eq:BichromaticHamJA}
\end{align}
\end{widetext}
where $J_n$ is the $n$th Bessel function of the first kind. By an appropriate choice of $\delta$, $\omega_g$, and motional frequencies, it is possible to ensure that only one of the terms in Eq.~(\ref{eq:BichromaticHamJA}) is near-resonant, while all the others oscillate many times over the gate duration and can thus be neglected. It is possible to have motional mode frequencies and values of $\delta$ such that multiple terms above are near resonance; depending on the parameters, these terms may or may not commute with each other, as some are proportional to $\hat{S}_{y}$ and others to $\hat{S}_{z}$. It is important to ensure that terms that do not commute with the dominant, near-resonant term are well off resonance at all times (including during the motional frequency ramping) to minimize their contributions, which cause uncorrectable gate errors. Off-resonant commuting terms simply rescale the rate at which geometric phase is accumulated, as long as their detunings from resonance remain sufficiently large during the gate that the associated spin-motion entanglement is adiabatically eliminated at the end of the gate interaction by the ramping processes (see next section). This eliminates the need to carefully calibrate the simultaneous closure of trajectories in multiple modes as is sometimes the case in gates without adiabatic ramping. It is important to choose motional frequencies, motional ramp parameters, and $\delta$ such that none of the terms in Eq.~(\ref{eq:BichromaticHamJA}) goes through resonance, or too close to resonance, during the frequency ramping process, as this violates some of our underlying assumptions and makes it difficult or impossible to eliminate spin-motion entanglement at the end of the gate, leading to errors. For this work, we choose the experimental parameters to select a single near-resonant term, the $J_2$ term for the out-of-phase mode in the out-of-plane radial direction; the frequency spectrum of farther-off-resonant terms is discussed in the main text. The single resonant term can be written as
\begin{widetext}
\begin{align}
    H_{\text{bi,single}} &\approx 4\cos(\omega_g t) \cos(2\delta t)\, \Omega_{g}(t) J_2\left(\frac{4\Omega_\mu(t)}{\delta}\right) ~\hat{S}_{z} \left(\hat{a} e^{-i\phi(t)} + \hat{a}\dagg e^{i\phi(t)}\right), \label{eq:SingleModeBichromaticHam}
\end{align}
\end{widetext}
where to simplify notation, we have dropped the $i$ and $c,s$ subscripts on the gradient Rabi rate, collective spin operator, and motional creation and annihiliation operators. We refer to the motional frequency of the mode as $\omega_m(t)=\omega_{m0}+\omega_{m1}\gamma(t)$, where $\omega_{m0}$ and $\omega_{m1}$ are constants, $\omega_{m0}>0$, $\omega_{m0}\gg|\omega_{m1}|$, and $\gamma(t)$ is the ramping function defined in Eq.~(\ref{eq:envelopeFunctions}). The accumulated motional phase is now ${\phi(t)=\intl_0^t \omega_m(t') dt'}$. The conditions for Eq.~(\ref{eq:SingleModeBichromaticHam}) to be the dominant, near-resonant term are that $\omega_g\pm2\delta\approx\omega_{m0}$, $\delta\gg|\omega_{m1}|$, ${\tau_\mu,\tau_g}\gg1/\delta$, and that the other motional mode frequencies are farther detuned at all times from $\omega_g\pm m\delta$ for all integers $m$. For our experiment, we have chosen $\omega_g+2\delta\approx\omega_{m0}$, and we can rewrite Eq.~(\ref{eq:SingleModeBichromaticHam}), eliminating fast-oscillating terms, as
\begin{align}
    H_\text{bi}(t) \equiv \Omega_{g}(t) J_2\left(\frac{4\Omega_\mu(t)}{\delta}\right) \hat{S}_z \left[\hat{a}\dagg e^{i\Phi(t)} + \hat{a} e^{-i\Phi(t)}\right]\, . \label{eq:HamMicroBi}
\end{align}
We have modified the accumulated phase to $\Phi(t) \equiv \int_0^t \Delta(t') dt' = \phi(t)-\omega_g t-2\delta t$, where $\Delta(t) \equiv \omega_{m}(t) - \omega_g - 2\delta$ is the detuning between the state-dependent force and the motional frequency. 

\section{Analytical model for infidelity of an amplitude and frequency ramped gate}

We now derive analytical expressions for the residual spin-motion entanglement at the end of a general geometric phase gate, showing how it can be adiabatically eliminated by ramping the amplitude of the state-dependent force along with its detuning from the motional frequency.  The general Hamiltonian for a detuned state-dependent force (SDF) acting on a set of $N$ qubits and driving state-dependent trajectories in phase space for a single motional mode can be written as

\begin{align}
    H_\text{gate}(t) \equiv \Omega_{\phi}f(t) \sum_{j=1}^Nc_j\hat{\sigma}^{(j)}_k \left[\hat{a}\dagg e^{i\Phi(t)} + \hat{a} e^{-i\Phi(t)}\right]\, , \label{eq:HamGeneral}
\end{align}
where $\Omega_\phi$ is a constant equal to the maximum strength of the SDF on any ion, $f(t)$ is a dimensionless function taking values in $[0,1]$ that captures the time dependence of the SDF amplitude, $\hat{a}$ and $\hat{a}\dagg$ are annihilation and creation operators for the motional mode, $\hat{\sigma}^{(j)}_k$ is a Pauli operator for the $j$th qubit with $k\in\{x,y,z\}$, the $c_j$ are products of the normalized  participation of the $j$th ion in the motional mode~\cite{James1998} with the relative SDF strength at the $j$th ion due to individually addressed or otherwise non-uniform control fields, and $\Phi(t) \equiv \int_0^t \Delta(t') dt'$. Here $\Delta(t)=\omega_m(t)-\omega_\text{SDF}(t)$ is the time-dependent detuning between the SDF frequency $\omega_\text{SDF}(t)$ and the motional mode frequency $\omega_m(t)$. As long as any time dependence of $\omega_m(t)$ is adiabatic with respect to its instantaneous value ($\left|\frac{1}{\omega_m^2}\frac{d\omega_m}{dt}\right|\ll1$) to avoid squeezing of the motional state, the time-dependent detuning $\Delta(t)$ can be realized by changing $\omega_m$, $\omega_{\text{SDF}}$, or both. We emphasize that Eq.~(\ref{eq:HamGeneral}) describes an arbitrary geometric phase gate on a single motional mode with a single SDF frequency, covering many gates that are currently used experimentally\footnote{We can map the specific gate performed in this work onto Eq.~(\ref{eq:HamGeneral}) by comparing with the approximate gate Hamiltonian in Eq.~(\ref{eq:HamMicroBi}), giving $\Omega_\phi=\sqrt{2}\Omega_{g0}J_2\left(\frac{4\Omega_{\mu0}}{\delta}\right)$ (the factor of $\sqrt{2}$ arises because the normalization convention for the $c_j$ in Eq.~(\ref{eq:HamGeneral}) with identical control fields at both ions differs by a factor of $1/\sqrt{2}$ relative to the definition of $\hat{S}_z$ in Eq.~(\ref{eq:HamMicroBi})), $k=z$, and $\Delta$ and $\Phi$ as defined for Eq.~(\ref{eq:HamMicroBi}), with fixed $\omega_{SDF}=\omega_g+2\delta$ and time-dependent $\omega_m(t)$. Because the microwave magnetic field amplitude $\Omega_\mu(t)$ that determines the SDF strength is inside the argument of a Bessel function, $f(t)$ does not have a simple analytical form for the specific instance of the gate in this work, but can be computed numerically from the envelope shape in Eq.~(\ref{eq:envelopeFunctions}), which for $\Omega_\mu(t)$ is a Blackman-Harris envelope~\cite{Harris1978}.}. The extension of the Hamiltonian to multiple motional modes and/or SDF frequencies is straightforward, and the subsequent derivation for the residual spin-motion entanglement can likewise be extended as long as its underlying assumptions are still met for all the motional mode and SDF frequencies. 

We divide the gate into three time regions: the first for $t\in[0,\tau]$ during which the SDF amplitude is ramped on from zero to its maximum value, the second for $t\in[\tau, t_f+\tau]$ when the detuning between the SDF and the motional frequency is ramped from far-detuned to near-detuned and back to far-detuned with fixed SDF amplitude, and the third for $t\in[t_f+\tau,t_f+2\tau]$ during which the SDF amplitude is ramped back to zero. The total gate duration is $T=t_f+2\tau$. 

The SDF amplitude follows a smooth arbitrary function $f(t)$ with vanishing first derivative at the start and end of the ramps. Specifically, each amplitude ramp is performed smoothly over a time $\tau$, with the boundary conditions $f(0) = f(t_f + 2\tau) = \dot f(0) = \dot f(t_f + 2\tau) = 0$, and with $f(t) = 1$ when $t\in[\tau, t_f+\tau]$.  We also assume that $f$ is continuous and differentiable everywhere, leading to $\dot f(\tau) = \dot f(t_f+\tau) = 0$. During the amplitude ``flat top'' regime, $t\in[\tau, t_f+\tau]$, the SDF-motion detuning $\Delta$ is smoothly ramped according to $\Delta(t) = \Delta_1 + \Delta_{r} \gamma(t)$, where $\Delta_1$ and $\Delta_{r}$ are positive constants with $\Delta_r\gg\Delta_1$, and $\gamma(t)$ is a dimensionless function taking values in $[0,1]$ that is symmetric in time around the midpoint $t=\tau+t_f/2$. We take $\gamma(t)= 1$, with vanishing first derivative, at the start and end of the ramp, writing $\gamma(t) = 1$ for $t\in[0,\tau]\cup[t_f+\tau,t_f+2\tau]$, $\dot \gamma(\tau) = \dot \gamma(t_f+\tau) = \dot \gamma(t_f/2+\tau) = 0$, and $\gamma(t_f/2+\tau) = 0$. We also assume that $\gamma$ is continuous and differentiable everywhere. Without loss of generality we have defined $\Delta>0$ at all times, but the derivation can equally well be carried out for $\Delta<0$ at all times ($\omega_\text{SDF}>\omega_m$), yielding the same scalings with the magnitude of $\Delta$; we note that in both cases, $\Delta$ must not change sign during the gate (the SDF should never become resonant with the motion).

Performing the Magnus expansion on Eq.~(\ref{eq:HamGeneral}) to second order (which is exact for this Hamiltonian) allows us to write the propagator for Eq.~(\ref{eq:HamGeneral}) as
\begin{align}
    U_{\text{gate}}(t) &= \exp\left[-i\left(\xi(t) \hat{a}\dagg + \xi^*(t) \hat{a}\right) \sum_{j=1}^Nc_j\hat{\sigma}^{(j)}_k\right] \exp(-i\theta(t) ~\sum_{j=1}^N\sum_{m=1}^Nc_jc_m\hat{\sigma}^{(j)}_k\hat{\sigma}^{(m)}_k),
\end{align}
where $\xi(t)$ is a complex-valued function describing the amplitude of the state-dependent displacement and $\theta(t)$ is a real-valued function expressing the geometric phase acquired. The first exponential is the propagator for a state dependent force that couples the motional mode and the spins, while the second exponential is the propagator for pairwise entanglement generation between the spins in a $\hat{\sigma}_k\otimes\hat{\sigma}_k$ gate interaction.  The ideal implementation of the gate has $\xi(T) = 0$ (no residual spin-motion entanglement at the end of the gate) and $\theta(T)$ chosen to give the desired geometric phase at the end of the gate interaction.  Focusing on the SDF, we find that the displacement obeys the integrable ODE
\begin{align}
    \dot\xi(t) &= \Omega_\phi f(t) e^{i\Phi(t)}.
\end{align}
We integrate this expression to determine the displacement $\xi(T)$ at the end of the gate. In the steps below, we perform integration by parts twice, after Eqs.~(\ref{eq:xiOriginal}) and (\ref{eq:intbyparts2}), and use the identity $e^{i\Phi(t)} = [i \dot\Phi(t)]^{-1}\dv{t} e^{i\Phi(t)}$ and the fact that $\dot\Phi(t) = \Delta(t)$. 
\begin{align}
    \xi(T) &= \Omega_\phi \intl_0^T f(t) e^{i\Phi(t)} dt \\
    &= \Omega_\phi\left[ \intl_0^\tau f(t) e^{i\Phi(t)} dt + \intl_\tau^{t_f+\tau}e^{i\Phi(t)} dt + \intl_{t_f+\tau}^{t_f + 2\tau} f(t) e^{i\Phi(t)} dt \right] \label{eq:xiOriginal}\\
    &= \Omega_\phi \Bigg[ \frac{f(t) e^{i\Phi(t)}}{i\dot\Phi(t)}\Bigg|_0^\tau + \frac{e^{i\Phi(t)}}{i\dot\Phi(t)}\Bigg|_\tau^{t_f + \tau} + \frac{f(t) e^{i\Phi(t)}}{i\dot\Phi(t)}\Bigg|_{t_f+\tau}^{t_f + 2\tau} - \intl_0^\tau \frac{\dot f(t) e^{i\Phi(t)}}{i\Delta_0}dt + \intl_\tau^{t_f + \tau} \frac{\dot\Delta(t) e^{i\Phi(t)}}{i\Delta^2(t)}dt - \intl_{t_f + \tau}^{t_f + 2\tau} \frac{\dot f(t) e^{i\Phi(t)}}{i\Delta_0}dt \Bigg] \label{eq:afterintbyparts1}\\
    &= \Omega_\phi \Bigg[- \intl_0^\tau \frac{\dot f(t) e^{i\Phi(t)}}{i\Delta_0}dt + \intl_\tau^{t_f + \tau} \frac{\dot\Delta(t) e^{i\Phi(t)}}{i\Delta^2(t)}dt - \intl_{t_f + \tau}^{t_f + 2\tau} \frac{\dot f(t) e^{i\Phi(t)}}{i\Delta_0}dt \Bigg] \label{eq:intbyparts2}\\
    &= \Omega_\phi \Bigg[- \intl_0^\tau \frac{\ddot f(t) e^{i\Phi(t)}}{\Delta_0^2}dt + \intl_\tau^{t_f + \tau} \left[\frac{\ddot\Delta(t) }{\Delta^3(t)} - \frac{3\dot\Delta^2(t)}{\Delta^4(t)}\right] e^{i\Phi(t)}dt - \intl_{t_f + \tau}^{t_f + 2\tau} \frac{\ddot f(t) e^{i\Phi(t)}}{\Delta_0^2}dt \Bigg] \equiv I_{\text{amp1}} + I_{\text{mot}} + I_{\text{amp2}}\label{eq:fullXi}
\end{align}
In going from (\ref{eq:intbyparts2}) to (\ref{eq:fullXi}) we used the fact that $\dot f(0) = \dot f(\tau) = \dot f(t_f + \tau) = \dot f(t_f+2\tau) = 0$, and $\dot\Delta(\tau) = \dot\Delta(t_f+\tau) = 0$.  We have defined the ``far detuning'' $\Delta_0\equiv\Delta(t\leq\tau) = \Delta(t\geq\tau+t_f)=\Delta_1+\Delta_{r}$ at the start and end of the frequency ramp and the ``near detuning'' $\Delta_1\equiv\Delta(\tau+t_f/2)$, and make the assumption that $\Delta_0\gg\Delta_1$, recalling that we have chosen the convention $\Delta>0$ at all times as described above. 

The first and last terms, $I_{\text{amp1}}$ and $I_{\text{amp2}}$, are of the form described in Ref.~\cite{Sutherland2024} for a gate with adiabatic amplitude ramping only, where it was shown that $I_{\text{amp1}} + I_{\text{amp2}} = \mathcal{O}(\Omega_\phi\Delta_0^{-3}\tau^{-2})$ and are hence adiabatically suppressed when $\tau\Delta_0\gg 1$ (slow ramping condition) and $\Omega_\phi/\Delta_0\ll 1$ (far detuning condition). The boundary terms from the frequency-ramped portion of the gate ($t\in[\tau,t_f+\tau]$), cancel exactly with the boundary terms of the amplitude ramps in Eq.~(\ref{eq:afterintbyparts1}) after the first integration by parts due to the construction of $f(t)$.  This is important as without proper cancellation, the first integration by parts would give boundary terms that do not vanish in the adiabatic limit.  Analogously, since $\gamma(t)$ is continuous and differentiable, any near-detuned ``flat-top'' region of the frequency ramp (where $\gamma(t)=0$ for $t\in[t_1, t_2]$, and thus $\dot{\gamma}(t_1)=\dot{\gamma}(t_2)=0$) will not contribute to the final displacement, as after the first integration by parts the boundary terms from the detuning flat-top region would cancel with the boundary terms from the neighboring detuning ramps. Ref.~\cite{Sutherland2024} shows that amplitude flat-top regions in amplitude-ramped gates exhibit the same cancellation and thus do not contribute to the final displacement $\xi(T)$. 

We emphasize that Eq.~(\ref{eq:fullXi}) is exact for any amplitude- and frequency-ramped geometric phase gate of the form in Eq.~(\ref{eq:HamGeneral}), including if the spin-motion interaction is $\hat{S}_x$ or $\hat{S}_y$ instead of $\hat{S}_z$, as long as $f(t)$ and $\gamma(t)$ (or equivalently $\Delta(t)$) satisfy the constraints described. Each additional order of zero derivative constraints on $f$, for example requiring $\ddot{f}(0)=\ddot{f}(\tau)=\ddot{f}(t_f+\tau)=\ddot{f}(t_f+2\tau)=0$, enables another round of integration by parts of the amplitude integrals with zero boundary terms, giving an additional power of $1/(\tau\Delta_0)$ to the scaling of $I_{\text{amp1}}$ and $I_{\text{amp2}}$. Additional zero derivative constraints on $\gamma$ produce more mathematically complicated expressions when integrating $I_\text{mot}$ by parts, but the same principle applies that each additional order of zero derivative of $\gamma$ at $t=\tau$ and $t=t_f+\tau$ produces an additional multiplicative factor of $1/(t_f\Delta)$ in the leading-order terms in $I_\text{mot}$. Thus we see the characteristic adiabatic tradeoff, where at the cost of time (because the ramps take longer when they start and end with more orders of zero time derivative), it is possible to further increase the adiabatic suppression of the final spin-motion entanglement. 

Eq.~(\ref{eq:fullXi}) can be solved numerically to optimize frequency or amplitude ramp shapes and durations. We can also find general analytic scaling laws for the middle integral $I_\text{mot}$, as we have for $I_{\text{amp1}}$ and $I_{\text{amp2}}$, to show how it is adiabatically suppressed. However, $I_\text{mot}$ is in the non-perturbative regime when $\Delta(t)$ is small or $\dot\Phi \approx 0$ (a stationary point), which happens in the region around $t = t_f/2+\tau$ when $\Delta(t_f/2+\tau) = \Delta_1$.  Near the stationary point, the exponential is oscillating slowly, and its contribution must be calculated through bulk integration via the stationary phase approximation.  We proceed by expanding $\Delta$ and $\Phi$ around this stationary point and then evaluating the integral through the stationary phase approximation.  We first change variables from $t$ to $s = (t-\tau)/t_f$, and then $u = s - 1/2$ (but noting that the dot notation in the expression below still means $\partial/\partial t$), giving
\begin{align}
    I_{\text{mot}} &= \Omega_\phi t_f \intl_0^1 \left[\frac{\ddot{\Delta}(s) }{\Delta^3(s)} - \frac{3\dot\Delta^2(s)}{\Delta^4(s)}\right] e^{i\Phi(s)}ds \label{eq:Imots} \\
    &\approx \Omega_\phi t_f e^{i\Delta_0\tau} e^{i\psi_0} \intl_{-1/2}^{1/2} \left[ \frac{\Delta_{r} \lambda/t_f^2}{[\Delta_1 + \frac{\Delta_{r}}{2}\lambda u^2]^3} - \frac{3\left(\Delta_{r} \lambda u/t_f \right)^2}{[\Delta_1 + \frac{\Delta_{r}}{2}\lambda u^2]^4} \right] e^{it_f (\Delta_1 u + \Delta_{r}/6 \lambda u^3)} du\\
    &\equiv \Omega_\phi t_f e^{i\Delta_0\tau} e^{i\psi_0} \intl_{-1/2}^{1/2} \left[ \frac{2x/t_f^3}{[\Delta_1 + \frac{x}{t_f} u^2]^3} - \frac{3\cdot (2x u/t_f^2)^2}{[\Delta_1 + \frac{x}{t_f} u^2]^4} \right] e^{i(t_f \Delta_1 u + x u^3/3)} du\\
    &\equiv \frac{2^{4/3} \Omega_\phi t_f^{2/3}}{\Delta_{r}^{1/3} \lambda{}^{1/3}} e^{i\Delta_0\tau} e^{i\psi_0} \intl_{-V}^{V} \left[ \frac{1}{[\eta + v^2]^3} - \frac{3 \cdot 2v^2}{[\eta + v^2]^4} \right] e^{i(\eta v + v^3/3)} dv \, . \label{eq:xiPreInfinity}
\end{align}
We have defined $\lambda = t_f^2\, \ddot\gamma(t_f/2+\tau)$, noting that $\lambda$ is a dimensionless constant of order unity. Because $\gamma$ is constructed with respect to the total frequency ramping duration $t_f$, it can be thought of as a function of $t/t_f$ rather than of $t$; therefore, each time derivative of $\gamma$ will accrue a factor of $1/t_f$, which we cancel by construction when defining $\lambda$. We also define $x = \Delta_{r} t_f \lambda/2$ and $\eta = \Delta_1 t_f/x^{1/3}$, and in Eq.~(\ref{eq:xiPreInfinity}) we made the change of variables $v = x^{1/3}u$ and $V = \frac12 x^{1/3}$.  We further define ${\psi(s)= t_f \int_0^s [\Delta_1 + \Delta_{r}\gamma(t_f s' + \tau)]ds'}$ and $\psi_0 = \psi(1/2)$.  If we work in the limit where $x$ is large ($\Delta_{r} t_f \gg 1$, roughly speaking the condition that the motional trajectory undergoes many loops in phase space over the course of the total frequency ramping waveform), then the integral will be dominated by the points, $s = s^*$, where $\dot\Phi(s^*)/t_f \approx 0$. We extend the integration limits in (\ref{eq:xiPreInfinity}) to $\infty$.  This is justified because we will be working in the large $V$ or $x$ limit, but will give rise to an error $\delta I_\infty$.  This error term scales as $\mathcal{O}(\Omega_\phi \Delta_0^{-3} t_f^{-2})$, as we derive later starting with Eq.~(\ref{eq:IVIntegral}). The integral can then be computed as
\begin{align}
    I_{\text{mot}} &= \frac{2^{4/3}\Omega_\phi t_f^{2/3}}{\Delta_{r}^{1/3}\lambda{}^{1/3}} e^{i\Delta_0\tau} e^{i\psi_0} \intl_{-\infty}^\infty \left[\frac{1}{(\eta + v^2)^3} - \frac{3\cdot 2 v^2}{(\eta + v^2)^4}\right] e^{i\left(v^3/3 + \eta v\right)} dv + \delta I_{\infty}\\
    &= \frac{2^{4/3}\Omega_\phi t_f^{2/3}}{\Delta_{r}^{1/3}\lambda{}^{1/3}} e^{i\Delta_0\tau} e^{i\psi_0} \intl_{-\infty}^\infty \dv{v}\left[\frac{v}{(\eta + v^2)^3}\right] e^{i\left(v^3/3 + \eta v\right)} dv + \mathcal{O}(\Omega_\phi \Delta_0^{-3} t_f^{-2})\\
    &= -\frac{2^{4/3}\Omega_\phi t_f^{2/3}}{\Delta_{r}^{1/3}\lambda{}^{1/3}} e^{i\Delta_0\tau} e^{i\psi_0} \intl_{-\infty}^\infty \frac{v}{(\eta + v^2)^3} \dv{v} e^{i\left(v^3/3 + \eta v\right)} dv + \mathcal{O}(\Omega_\phi \Delta_0^{-3} t_f^{-2})\\
    &= \frac{2^{1/3}\Omega_\phi t_f^{2/3}}{\Delta_{r}^{1/3}\lambda{}^{1/3}} e^{i\Delta_0\tau} e^{i\psi_0} \intl_{-\infty}^\infty e^{i\left(v^3/3 + \eta v\right)} dv + \mathcal{O}(\Omega_\phi \Delta_0^{-3} t_f^{-2}).
\end{align}
We can now proceed by using the complex integral definition of the Airy function $\airy(\eta)$, and adding together all the contributions to the final displacement we find
\begin{align}
    \xi(T) &= I_\text{mot} + I_{\text{amp1}} + I_{\text{amp2}} \\
    &= \frac{2^{4/3} \pi \Omega_\phi t_f^{2/3}}{\Delta_{r}^{1/3}\lambda{}^{1/3}} e^{i\Delta_0\tau} e^{i\psi_0} \airy(\eta) + \mathcal{O}(\Omega_\phi \Delta_0^{-3} t_f^{-2}) + \mathcal{O}(\Omega_\phi \Delta_0^{-3}\tau^{-2}).
\end{align}
Making use of the asymptotic expansion for an Airy function, valid for $\eta\gtrsim1$ \cite{NIST:DLMF},
\begin{align}
    \airy(\eta) &\sim \frac{1}{2\sqrt{\pi\sqrt{\eta}}}\exp(-\frac23 \eta^{3/2}) \\
    \implies \xi(T) &= 2^{1/4}\sqrt\pi \Omega_\phi e^{i\Delta_0\tau} e^{i\psi_0} \sqrt{\frac{t_f}{\sqrt{\Delta_{r} \Delta_1 \lambda}}} \exp(-\frac{2\sqrt2 \Delta_1^{3/2}}{3\sqrt{\lambda \Delta_{r}}} t_f) + \mathcal{O}(\Omega_\phi \Delta_0^{-3} t_f^{-2}) + \mathcal{O}(\Omega_\phi \Delta_0^{-3}\tau^{-2})\\
    &\approx \left(\frac{2\pi^2}{\lambda}\right)^{1/4} \Omega_\phi e^{i\Delta_0\tau} e^{i\psi_0} \sqrt{\frac{t_f}{\sqrt{\Delta_0 \Delta_1}}} \exp(-\sqrt\frac{8}{9\lambda} \frac{\Delta_1^{3/2}t_f}{\Delta_0^{1/2}}) + \mathcal{O}(\Omega_\phi \Delta_0^{-3} t_f^{-2}) + \mathcal{O}(\Omega_\phi \Delta_0^{-3}\tau^{-2}) \, ,\label{eq:xiFinal}
\end{align}
where we have used the fact that $\Delta_0\gg\Delta_1$ to make the approximation $\Delta_{r}\approx\Delta_0$ in the final step. The motional frequency ramp contribution has a term that is exponentially suppressed in $t_f$ (requiring ${t_f\Delta_1^{3/2}\Delta_0^{-1/2}\gg1}$) and a term that is suppressed as $\mathcal{O}(\Omega_\phi \Delta_0^{-3}t_f^{-2})$, while the amplitude ramp contribution is dominated by a term that goes as $\mathcal{O}(\Omega_\phi \Delta_0^{-3}\tau^{-2})$. The first term (the exponential) sets a practical limit on how small the near detuning $\Delta_1$ can be, providing a tradeoff between the increased gate speed achievable with smaller $\Delta_1$ and increased residual state-dependent displacement at the end of the gate. 

As mentioned earlier, extra constraints that force additional higher derivatives of $\gamma$ ($\ddot{\gamma}, \dddot{\gamma}$, etc.) to zero at $t=\tau$, $t=t_f/2+\tau$, and $t=t_f+\tau$ will alter these scalings. Starting from Eq.~(\ref{eq:Imots}), it is possible to derive equivalent expressions to Eq.~(\ref{eq:xiFinal}) using the same method but with additional rounds of integration by parts. It can be shown that the exponential term in Eq.~(\ref{eq:xiFinal}) has the same scaling in this instance, but the term due to $\delta I_\infty$ will have an additional factor of $\Delta_0^{-1} t_f^{-1}$ for each additional order of zero derivative in $\gamma$. 

To quantify how the residual spin-dependent displacement contributes to gate infidelity, we assume that the state starts in a motional Fock state $\ket{n}$ and then compute the final infidelity with respect to this same Fock state, averaging over all initial pure spin states in $\text{SU}(4)$. In the regime where $|\xi(T)|\ll1$, then the infidelity is
\begin{align}
    \mathcal{I} &\approx \abs{\xi(T)}^2 (2n+1) \lambda_{S_z}^2 \\
    &= \frac85 \abs{\xi(T)}^2 (2n+1) \, ,
\end{align}
where $\lambda_{S_z}^2$ is the variance of $\hat{S}_z$ and is equal to $8/5$ when averaged over $\text{SU}(4)$ (an average over all initial pure states).  Hence, the infidelity will have terms that scale as the motional occupation times $\mathcal{O}(\Omega_\phi^2 \Delta_0^{-6}t_f^{-4})$, $\mathcal{O}(\Omega_\phi^2 \Delta_0^{-6}t_f^{-2}\tau^{-2})$, and $\mathcal{O}(\Omega_\phi^2 \Delta_0^{-6}\tau^{-4})$, meaning that sensitivity to the motional state is substantially reduced when working at large far detunings and long ramp durations.  We have assumed that we are working with sufficiently large near detuning $\Delta_1$ and $t_f$ so that the the contribution of the first term in Eq.~(\ref{eq:xiFinal}) is exponentially small and can be neglected. For the specific case of the gate performed in this work $\Omega_\phi = \mathcal{O}(\Omega_{g0})$ (see footnote earlier), so we have made this substitution into the expressions above where they appear in the End Matter. 

The relative insensitivity of this infidelity to offsets or miscalibrations in the motional frequency can be seen from these scalings as well. As long as $\Delta_1$ and $t_f$ are large enough to be able to neglect the exponential term in Eq.~(\ref{eq:xiFinal}), the impact of an offset in the motional frequency will enter as a correction to $\Delta_0$. However, since $\Delta_0$ is by design very large, this will be a small fractional correction.  With sufficient adiabatic suppression, the primary source of infidelity from motional frequency offsets or miscalibrations is due to failure to accumulate the correct geometric phase $\theta(T)$, rather than failure to eliminate residual spin-motion entanglement at the end of the gate. 
\subsection{Error from extending integration limits}
In the previous section, we extended the integration limits of the expression for $\xi(T)$ to infinity, so that the integral could be calculated analytically.  This process contributes a dominant term of $\mathcal{O}(\Omega_\phi \Delta_0^{-3} t_f^{-2})$, which we now derive.  Our starting point is the integral in (\ref{eq:xiPreInfinity}),
\begin{align}
    I(V) &= \intl_{-V}^V \left[\frac{1}{(\eta + v^2)^3} - \frac{3\cdot 2 v^2}{(\eta + v^2)^4}\right] e^{i\left(v^3/3 + \eta v\right)} dv \label{eq:IVIntegral}.
\end{align}
with $V = \frac12 x^{1/3}$.  However, the second term in the integrand, prior to the changes of variable, is due to a ratio given by $\dot\Delta^2/\Delta^4$.  Since this term is polynomial in $\dot\Delta$, it can be integrated by parts repeatedly prior to a change of variables, and any resulting terms proportional to $\dot\Delta$ will vanish on the boundary.  The end result is a term that is subleading when compared to the first term in the integrand in (\ref{eq:IVIntegral}).  Hence, we focus on the first term and write the integral as,
\begin{align}
    I(V) &\approx \intl_{-V}^V \frac{e^{i\left(v^3/3 + \eta v\right)}}{(\eta + v^2)^3} dv \\
    &= I(\infty) - I(|v| > V).
\end{align}
Our task is to quantify $I(|v|>V)$.  We proceed through integration by parts,  where the boundary term dominates as the next term goes as $\dddot{\Delta}/\Delta^5$:
\begin{align}
    I(|v|>V) &= \frac{e^{i\left(v^3/3 + \eta v\right)}}{i(\eta + v^2)^4} \Bigg|_V^{\infty} + \frac{e^{i\left(v^3/3 + \eta v\right)}}{i(\eta + v^2)^4} \Bigg|_{-\infty}^{-V} + \intl_{|v| > V} \frac{4\dddot\Delta}{i\Delta^5} e^{i\Phi} \\
    &\approx -2\frac{\sin(V^3/3 + \eta V)}{V^8} \\ 
    \implies I(|v|>V) &= \mathcal{O}(x^{-8/3}) \\
    &= \mathcal{O}(\Delta_{r}^{-8/3} t_f^{-8/3}).
\end{align}
Combining this result with the prefactor in (\ref{eq:xiPreInfinity}), we find that extending the integral to infinity neglected a term that is $\mathcal{O}(\Omega_\phi \Delta_{r}^{-3} t_f^{-2})$.  With our assumption of large far detuning, $\Delta_0\approx\Delta_{r}$, this integral correction is approximately
\begin{align}
    \delta I_{\infty} &= \mathcal{O}(\Omega_\phi \Delta_0^{-3}t_f^{-2}).
\end{align}

\bibliographystyle{apsrev4-2}
\providecommand{\noopsort}[1]{}\providecommand{\singleletter}[1]{#1}%